\documentstyle[pra,aps,multicol,eqsecnum,epsfig]{revtex} 

\def\be{\begin{eqnarray}}
\def\ee{\end{eqnarray}}
\def\l{\langle}
\def\r{\rangle}   
\def\bo{\begin{figure}}
\def\eo{\end{figure}}

\begin{document}
\title{Probabilistic implementation of universal quantum processors}
\author{Mark Hillery${}^{1}$,
Vladim\'{\i}r Bu\v{z}ek${}^{2,3}$,
and M\'ario Ziman${}^{2}$}
\address{
${}^{1}$Department of Physics and Astronomy, Hunter College of CUNY, 695,
Park
Avenue, New York, NY 10021, U.S.A.\\
${}^{2}$Institute of Physics, Slovak Academy of Sciences,
D\'ubravsk\'a cesta 9, 842 28 Bratislava, Slovakia \\
${}^{3}$Faculty of Informatics, Masaryk University, Botanick\'a 68a,
602 00 Brno, Czech Republic 
}  

\date{25 May 2001}    

\maketitle

\begin{abstract}
We  present a probabilistic quantum processor for qudits.
The processor itself  is represented by a fixed array of gates. The input
of the processor consists of  two registers.
In the program register the set of instructions (program) is
encoded. This program is applied to the data register.
The processor 
 can perform any operation on a single qudit of the dimension $N$ 
with a certain probability.  If the operation is unitary, the
probability is in general $1/N^2$, but for more restricted sets
of operators the probability can be higher. In fact, this probability
can be independent of the dimension of the qudit Hilbert space of the
qudit under some conditions.
\end{abstract}


\begin{multicols}{2}

\section{Introduction}
\label{sec1}

Schematically we can represent a classical  computer
as a device with a processor, which is a fixed piece of hardware,
that performs operations on a {\em data} register according to a 
program encoded initially in the {\em program} register. The action 
of the processor is fully determined by the program. The processor 
is universal if we can realize any operation on the data by entering
the appropriate program into the program register.

In this paper we shall examine a quantum version of this picture.
Specifically, in close analogy with recent papers by Nielsen and
Chuang \cite{Nielsen1997} and Vidal and Cirac \cite{Vidal2000}, we will 
study how a quantum program initially put into a program register can
cause a particular operation to be applied to a data register initially prepared in an unknown  state.  We shall first consider the case in
which the data consists of a single qubit, and the program of two
qubits.  We shall then examine higher-dimensional systems.

Nielsen and Chuang  \cite{Nielsen1997}  originally formulated the
problem  in terms of a programmable array of quantum gates, which 
can be described as a fixed unitary operator, ${P}_{dp}$, that 
acts on both the program and the data.  The initial state, 
$|\Xi_{U}\r_p$, of the program register stores
information about the one-qubit unitary transformation ${U}$ that is 
going to be performed on a single-qubit data register initially prepared
in a state $|\psi\r_d$. The total dynamics of the programmable quantum
gate array is then given by
\be
{P}_{dp} \left[ |\psi\r_d\otimes |\Xi_U\r_p\right] = \left(U|\psi\r_d\right)
\otimes |\tilde{\Xi}_U\r_p,
\label{1.1}
\ee  
where only pure data states were considered. The program register at the
output of the gate is in the state $|\tilde{\Xi}_U\r_p$ - which was shown
to be independent of the input data state $|\psi\r_d$. 

Nielsen and Chuang proved that
any two inequivalent operations $U$ and $V$ require orthogonal program
states, i.e. $\l\Xi_U|\Xi_V\r=0$. Thus, in order to perfectly
implement a set of inequivalent operations,$\{U_j |j\in J\}$, 
the state space for the program register must contain the 
orthonormal set of program states, $\{|\Xi_{U_{j}}\r |j\in J\}$.
This means that the dimension of the program register must be
at least as great as the number of unitary operators that we
want to perform. 
Since the set of unitary operations is infinite, the result 
of Nielsen and Chuang implies that
no universal gate array can be constructed using finite resources, 
that is, with a finite dimensional program register.  They
did show, however, that if the gate array is probabilistic, a
universal gate array is possible.  A probabilistic array is one
that requires a measurement to be made at the output of the 
program register, and the output of the data register is only
accepted if a particular result, or set of results, is obtained.
This will happen with a probability, which is less than one.

Vidal and Cirac \cite{Vidal2000} have recently presented 
a probabilistic programmable quantum gate array with a finite
program register, which can realize a one parameter family
of operations, where the parameter is continuous, with
arbitrarily high probability.  The higher the probability of
success, the greater the dimensionality of the 
program register, but the
number of transformations that can be realized is infinite.
They have also considered {\em approximate} programmable 
quantum gate arrays, which perform an operation $E_U$ very 
similar to the desired $U$, that is $F(E_U,U) \geq 1-\epsilon$ for 
some transformation fidelity $F$.                           

Another aspect of the encoding of quantum operations in the states of
program registers has been discussed by Huelga and coworkers 
\cite{Huelga2000}. In this paper the implementation of an arbitrary
unitary operation $U$ upon a distant quantum system has been considered.
This so called teleportation of unitary operations has been formally
represented as a completely positive, linear,
trace preserving map on the set of density operators of the program
and data registers:
\be
{\cal T}\left[|\xi\r_{ab} \otimes |\Xi_U\r_p\otimes|\psi\r_d
\right]=
|\tilde{\xi}_{U}\r_{ap}\otimes \left(U|\psi\r_d\right)
\label{1.2}
\ee
Here $|\xi\r_a$  represents a specific entangled
state that is shared by two parties, Alice and Bob, who want to 
teleport the unitary operation $U$ from Alice to Bob. Huelga et al. 
\cite{Huelga2000} have investigated protocols which achieve the
teleportation of $U$ using local operations, classical communication
and shared entanglement. 

In the present paper we will address the problem of implementing an
operation $U$, encoded in the state of a program register 
$|\Xi_U\r_p$,  on the  data state $|\psi\r_d$. The gate arrays
we present are probabilistic; the program register must be measured
at the end of the procedure.
In Section \ref{sec2} we present a simple example of how to
apply an arbitrary operation to a single qubit initially prepared
in a state $|\psi\r$. The gate array consists of four Controlled-NOT
(C-NOT) gates, and can implement four programs perfectly.  These programs
cause the one of the operations $\openone$, $\sigma_{x}$, $-i\sigma_{y}$, 
or $\sigma_{z}$ to be performed on the data qubit.  Here $\openone$ is the
identity and $\sigma_{j}$, where $j=x,y,z$ is a Pauli matrix.  By
choosing programs that are linear combinations of the four basic ones,
it is possible to probabilistically perform any linear operation on 
the data qubit.  In Section \ref{sec3} we generalize the idea
to an arbitrary dimensional quantum system, a qudit.

\section{Operations on qubits}
\label{sec2}
We would like to construct a device that will do the following:
The input consists of a qubit, $|\psi\rangle_d$, and a second
state, $|\Xi_U\rangle_p$, which may be a multiqubit state, that
acts as a program.  The output of the device will be a state
$U|\psi\rangle_d$, where $U$ is an  operation that is specified
by $|\Xi_U\rangle_p$.  In order to make this a little less abstract,
we first  consider an example: 
Let $|\phi\rangle$ and $|\phi_{\perp}\rangle$ be two orthogonal 
qubit states, and suppose that we want to perform the operation
\begin{eqnarray}
\label{2.1}
A_{z} =  |\phi_{\perp}\rangle\langle\phi_{\perp}|-|\phi\rangle
\langle\phi | 
  =  \openone -2|\phi\rangle\langle\phi | ,
\end{eqnarray}
on $|\psi\rangle_d$.  The action of this operator is analogous to
that of $\sigma_{z}$ in the basis $\{ |0\rangle , |1\rangle \}$,
except that it acts in the basis $\{ |\phi_{\perp}\rangle , 
|\phi\rangle \}$.  That is, $\sigma_{z}$ does nothing to
$|0\rangle$ and multiplies $|1\rangle$ by $-1$, while $A_{z}$
does nothing to $|\phi_{\perp}\rangle$ and multiplies 
$|\phi\rangle$ by $-1$.  Can we  find a network and a program 
vector to implement this operation on $|\psi\rangle_d$?

We can, in fact, do this by using the network for a 
quantum information distributor (QID) as introduced in
Ref.~\cite{Braunstein2000} (this is a modification of
the quantum cloning transformation \cite{Buzek1996,Buzek1997}
).
In this network the program
register is represented by a two qubit state
$|\Xi_A\rangle_p$.  Before we present the network for the
programmable gate array, we shall introduce notation for 
its components. A Controlled-NOT gate $D_{jk}$ acting on 
qubits $j$ and $k$ performs the transformation,
\begin{equation}
D_{jk}|m\rangle_{j}|n\rangle_{k}=|m\rangle_{j}|m\oplus n\rangle_{k} ,
\label{2.2}
\end{equation}
where $j$ is the control bit, $k$ is the target bit, and $m$ 
and $n$ are either $0$ or $1$.  The addition is modulo $2$.
The QID network consists of four Controlled-NOT gates, and
acts on three qubits (a single data qubit denoted by a subscript 1 
and two program qubits denoted by subscripts 2 and 3, respectively).
Its action is given by the operator $P_{123}=D_{31}D_{21}D_{13}D_{12}$. 
As our first task, we shall determine how this network acts on
input states where qubit $1$ is in the state $|\psi\rangle$,
and qubits $2$ and $3$ are in Bell basis states.  The Bell
basis states are defined by
\begin{eqnarray}
\label{2.3}
|\Phi_{+}\rangle & = &\frac{1}{\sqrt{2}}(|01\rangle 
+ |10\rangle )  \equiv |\Xi_{01}\r \, ,  \nonumber \\
|\Phi_{-}\rangle & = &\frac{1}{\sqrt{2}}(|01\rangle 
- |10\rangle )  \equiv |\Xi_{11}\r \, ; \nonumber \\
|\Psi_{+}\rangle & = &\frac{1}{\sqrt{2}}(|00\rangle 
+ |11\rangle )\equiv |\Xi_{00}\r \, ;  \nonumber \\
|\Psi_{-}\rangle & = &\frac{1}{\sqrt{2}}(|00\rangle 
- |11\rangle )\equiv |\Xi_{10}\r \, . 
\end{eqnarray}
We find that 
\begin{eqnarray}
\label{2.4}
P_{123}|\psi\rangle_{1}|\Phi_{+}\rangle_{23} & = &(\sigma_{x}|\psi
\rangle_{1})|\Phi_{+}\rangle \, ;
\nonumber \\
P_{123}|\psi\rangle_{1}|\Phi_{-}\rangle_{23} & = & (-i\sigma_{y}
|\psi\rangle_{1})|\Phi_{-}\rangle \, ; 
\nonumber \\
P_{123}|\psi\rangle_{1}|\Psi_{+}\rangle_{23} & = & |\psi\rangle_{1}
|\Psi_{+}\rangle \, ; 
\nonumber \\
P_{123}|\psi\rangle_{1}|\Psi_{-}\rangle_{23} &=& (\sigma_{z}|\psi
\rangle_{1})|\Psi_{-}\rangle  .
\end{eqnarray}

Any operation on qubits can be expanded in terms of Pauli matrixes
and the identity.  The above equations mean that the Bell basis
vectors are ``programs'' for a complete set of operations.  In
order to see how to make use of this, let us expand our proposed
operation in terms of this complete set.  Expressing $|\phi\rangle$
as $|\phi\rangle = \mu |0\rangle +\nu |1\rangle$, we have that
\begin{eqnarray}
\label{2.5}
A_z & = & \openone -2|\phi\rangle\langle\phi | =\left(\begin{array}{cc} 
|\nu |^{2}-|\mu |^{2} & -2\mu\nu^{\ast} \\ -2\mu^{\ast}\nu 
& |\mu |^{2}-|\nu |^{2} \end{array}\right) ,  \\
 & = & -(\mu\nu^{\ast}+\mu^{\ast}\nu )\sigma_{x}+(\mu\nu^{\ast}
-\mu^{\ast}\nu )(-i\sigma_{y})
\nonumber
\\
& &+(|\nu |^{2}-|\mu |^{2})\sigma_{z} .
\nonumber
\end{eqnarray}
We can now apply the operation $A$ to $|\psi\rangle$ by sending
in the ``program'' vector
\be
\label{2.6}
|\Xi_A\rangle_{23}= & - & (\mu\nu^{\ast}+\mu^{\ast}\nu )|\Phi_{+}\rangle_{23}
+(\mu\nu^{\ast}-\mu^{\ast}\nu )|\Phi_{-}\rangle_{23}
\nonumber \\
& + & (|\nu |^{2}-|\mu |^{2})|\Psi_{-}\rangle_{23} ,
\ee
and measuring the program outputs in order to determine if they
are in the state $(|\Phi_{+}\rangle + |\Phi_{-}\rangle +|\Psi_{-}
\rangle) /\sqrt{3}$.  If they are, our operation has been accomplished.
Note that the measurement is independent of the vector 
$|\phi\rangle$ so that no knowledge of this vector is necessary
to make the measurement and to determine whether the procedure
has been successful. As we see, the probability of success is 1/3
for the implementation of the operation $A_z$ which is parameterized in
general by two continuous parameters (i.e. the state $|\phi\r$).

Let us examine the program vector more carefully.  If we define the
unitary operation, $U_{init}$, by
\begin{eqnarray}
\label{2.7}
U_{init}|00\rangle & = & -|10\rangle\, ; \qquad U_{init}|10\rangle
= -|11\rangle\, ; \nonumber \\
U_{init}|11\rangle & = & |01\rangle \, ; \qquad
U_{init}|01\rangle =  |00\rangle\, ; 
\end{eqnarray}
we have that
\begin{equation}
|\Xi_A \rangle_{12} = U_{init}\frac{1}{\sqrt{2}}(|\phi\rangle |\phi_{\perp}
\rangle +|\phi_{\perp}\rangle |\phi\rangle ).
\end{equation} 

Finally, we can summarize our procedure.  The steps are 
\begin{description}
\item[1.] Start with the state $\frac{1}{\sqrt{2}}(|\phi\rangle 
|\phi_{\perp} \rangle +|\phi_{\perp}\rangle |\phi\rangle )$ .
\item[2.] Apply $U_{init}$. 
\item[3.] Send the resulting state into the control ports (inputs $2$ and 
$3$) and $|\psi\rangle$ into port 1. 
\item[4.] Measure $(|\Phi_{+}\rangle + |\Phi_{-}\rangle +|\Psi_{-}
\rangle )/\sqrt{3}$ at the output of the control ports. 
\item[5.] If the result is yes, then the output of port $1$ is $(\openone -
2|\phi\rangle\langle\phi |)|\psi\rangle$.
\end{description}

Before proceeding to a more general consideration of this network,
let us make an observation.  Suppose that
we carry out the same procedure, but instead of starting with the 
program vector $(|\phi\rangle |\phi_{\perp} \rangle 
+|\phi_{\perp}\rangle |\phi\rangle )/\sqrt{2}$, we start instead with
the program vector $(|\phi\rangle |\phi\rangle -|\phi_{\perp}\rangle 
|\phi_{\perp}\rangle  )/\sqrt{2}$.  At the end of the procedure 
the output of the data register is $A_{x}|\psi\rangle$, where
\begin{equation}
A_{x}=|\phi\rangle\langle\phi_{\perp}|+|\phi_{\perp}\rangle\langle
\phi | .
\end{equation}
The operation $A_{x}$ interchanges $|\phi\rangle$ and $|\phi_{\perp}
\rangle$.  Its action is analogous to that of $\sigma_{x}$, which
interchanges the vectors $|0\rangle$ and $|1\rangle$.  The probability 
of success for this procedure is also $1/3$.

We now need to determine whether there is a program for any operator 
that could act on $|\psi\rangle$.  The operator need not be unitary;
it could be a result of coupling $|\psi\rangle$ to an ancilla,
evolving the coupled system (a unitary process), and then measuring 
the ancilla.  Therefore, if $A$ is now any linear operator acting
on a two dimensional quantum system, the transformations in which 
we are interested are given by
\begin{equation}
\label{trans}
|\psi\rangle \rightarrow \frac{1}{\| A\psi\|}A|\psi\rangle .
\end{equation}
Let us denote the operators, which can be implemented by Bell
state programs, by $S_{00}=\openone$, $S_{01}=\sigma_{x}$, $S_{10}=
\sigma_{z}$, and $S_{11}=-i\sigma_{y}$.  Any $2\times 2$ matrix
can be expanded in terms of these operators, so that we have
\begin{equation}
A=\sum_{j,k=0}^{1}\tilde{a}_{jk}S_{jk} .
\end{equation}
We now define $a_{jk}=\tilde{a}_{jk}/\sqrt{\eta}$, where
\begin{equation}
\eta = \sum_{j,k=0}^{1}|\tilde{a}_{jk}|^{2} ,
\end{equation}
so that
\begin{equation}
1 = \sum_{j,k=0}^{1}|a_{jk}|^{2} .
\end{equation}

Now let us go back to our network and consider the program vector 
given by 
\begin{equation}
|\Xi_{A}\rangle = \sum_{j,k=0}^{1}a_{jk}|\Xi_{jk}\rangle ,
\end{equation}
and at the output of the program register we shall measure the
projection operator corresponding to the vector $(1/2)
\sum_{j,k=0}^{1}|\Xi_{jk}\rangle$.  If the measurement is
successful, the state of the data register is, up to normalization,
given by
\begin{equation}
|\psi\r \rightarrow \left( \sum_{j,k=0}^{1}a_{jk}S_{jk}\right)
|\psi\r \, .
\end{equation}
After this state is normalized, it is just $(1/\| A\psi\| )
|\psi\rangle$.  This means that for any transformation of the type
given in Eq. (\ref{trans}), we can find a program for our network
that will carry it out.

\section{Generalization to qudits}
\label{sec3}
In order to extend the network presented in the previous section to
higher dimensions, we must first introduce a generalization of the 
two-qubit C-NOT gate \cite{Braunstein2000}
(see also Ref.~\cite{Alber2000}).  
As we noted previously, it is possible express the action of a 
C-NOT gate as a two-qubit operator of the form
\begin{eqnarray}
\label{3.1}
{D}_{ab}=\sum_{k,m=0}^1 |k\rangle_a\langle k|\otimes
|m\oplus k \rangle_b\langle m|\;. 
\end{eqnarray}       
In principle one can also introduce an operator 
${D}^\dagger_{ab}$ defined as
\begin{eqnarray}
\label{3.2}
{D}^\dagger_{ab}=\sum_{k,m=0}^1 |k\rangle_a\langle k|\otimes
|m \ominus k\rangle_b\langle m|\;. 
\end{eqnarray}       
In the case of qubits these two operators are equal, but this will
not be the case when we generalize the operator to Hilbert spaces
whose dimension is larger than 2 \cite{Braunstein2000,Alber2000}. 
In particular, we can generalize the operator ${D}$ for dimension
$N>2$ by defining   
\begin{eqnarray}
\label{3.3}
{D}_{ab}=
\sum_{k,m=0}^{N-1} |k\rangle_a\langle k|\otimes
|(m+k){\rm mod}\,N\rangle_b\langle m|\; , 
\end{eqnarray}
which implies that 
\begin{eqnarray}
{D}_{ab}^{\dagger}=
\sum_{k,m=0}^{N-1} |k\rangle_a\langle k|\otimes
|(m-k){\rm mod}\,N\rangle_b\langle m|\; . 
\end{eqnarray}       
From this definition it follows that the operator
${D}_{ab}$ acts on the basis vectors as
\be
{D}_{ab}|k\r|m\r = |k\r|(k+m) {\rm mod}\, N\r\;,
\label{3.4}
\ee
which means that this operator has the same action as
the conditional adder and can be performed with the help of
the simple quantum network discussed in \cite{Vedral1996}.
Now we see that for $N>2$ 
the two operators $ D$ and ${D}^\dagger$ do differ;
they describe conditional shifts in opposite directions.
Therefore the generalizations of the C-NOT operator to higher 
dimensions are just {\em conditional shifts}. 

In analogy with the quantum computational network discussed in the
previous section, 
we assume the network for the probabilistic universal quantum
processor to be
\be
{P}_{123}=
{D}_{31}{D}_{21}^\dagger{D}_{13}{D}_{12}\; .
\label{3.5}
\ee
The data register consists of system $1$ and the program register of
systems $2$ and $3$.  The state $|\Xi_U\rangle_{23}$ acts as  
the ``software'' which the operation to be implemented on the 
qudit data state $|\Psi\r_1$.  The output state 
of the three qudit system, after the four controlled 
shifts are applied, reads 
\begin{eqnarray} 
\label{3.6}
|\Omega\rangle_{123}= 
{D}_{31}{D}_{21}^\dagger{D}_{13}{D}_{12}
|\Psi\rangle_1|\Xi_U\rangle_{23}\;.
\end{eqnarray}     
A graphical representation of the logical network
(\ref{3.6}) with the conditional shift gates ${D}_{ab}$
in Fig.~\ref{fig1}.

\bo
\centerline{\epsfig{width=8.3cm, file=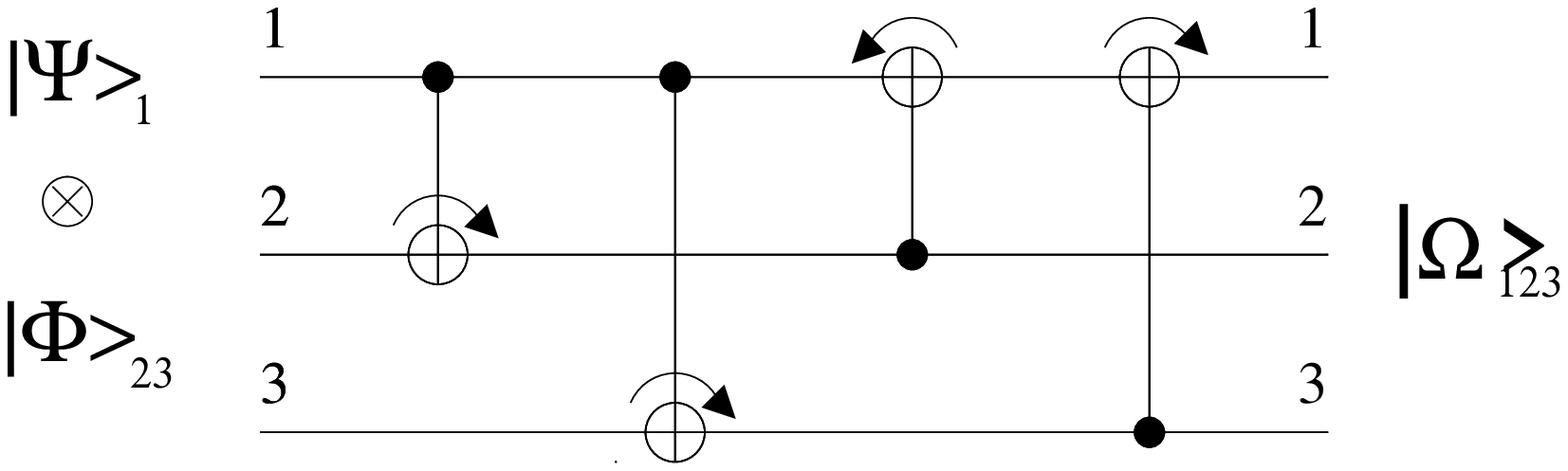}}
\bigskip
\caption{
A logic network for the universal quantum processor 
 as given by the unitary transformation
(\ref{3.6}). 
The action of the controlled 
shift operator $D_{jk}$ is represented as follows:
The control qudit is represented by {\large $\bullet$} while
the target qudit is represented { $\oplus$} with the rightarrow.
The action of the operator $D_{jk}^\dagger$ is represented by left arrow.
}
\label{fig1}
\eo      
The sequence of four operators acting on the basis vectors
gives 
$|n\rangle_{1}|m\rangle_{2}|k\rangle_{3}$ as
\end{multicols}
\vspace{-0.2cm}
\noindent\rule{0.5\textwidth}{0.4pt}\rule{0.4pt}{0.6\baselineskip}
\vspace{0.2cm}     
\begin{eqnarray} 
\label{3.7}
{D}_{31}{D}_{21}^\dagger{D}_{13}{D}_{12}
|n\rangle_{1}|m\rangle_{2}|k\rangle_{3}=
|(n-m+k){\rm mod}\,N\rangle_{1}\,|(m+n){\rm mod}\,N\rangle_{2}
\,|(k+n){\rm mod}\,N\rangle_{3}\;.
\end{eqnarray}     
  \hfill\noindent\rule[-0.6\baselineskip]%
  {0.4pt}{0.6\baselineskip}\rule{0.5\textwidth}{0.4pt}
\vspace{-0.2cm}
\begin{multicols}{2} 
We now turn to the fundamental program states.
A basis consisting of maximally entangled two-particle states 
(the analogue of the Bell basis for spin-$\frac{1}{2}$ particles) 
is given by
\cite{Fivel1995}
\begin{eqnarray}
\label{3.8}
|\Xi_{mn}\rangle = \frac{1}{\sqrt{N}} \sum_{k=0}^{N-1} \exp \Bigl(
i\frac{2\pi}{N} mk \Bigr) |k\rangle|(k-n){\rm mod}\,N\rangle \,,\!\!\!
\end{eqnarray}
where $m,n=0,\dots,N-1$.  If $|\Xi_{mn}\rangle_p$ is the initial 
state of the program register,
and $|\Psi\rangle=\sum_j \alpha_j |j\rangle_d$
(here, as usual,  $\sum_j |\alpha_j|^2=1$) is the initial state of
the data register, it then follows that
\be
&& {P}_{123}|\Psi\rangle_{1}|\Xi_{mn}\rangle_{23} = 
\nonumber\\
&=&
\sum_{jk}\frac{\alpha_j}{\sqrt{N}}\exp\frac{2\pi ikm}{N}{P}_{123}
|j\rangle |k\rangle |k-n\rangle
\nonumber\\
&=&\sum_{jk}\frac{\alpha_j}{\sqrt{N}}\exp\frac{2\pi ikm}{N}
|j-n\rangle |k+j\rangle |k+j-n\rangle
\nonumber\\
&=&\sum_{jk}\alpha_j\exp\frac{-2\pi ijm}{N}|j-n\rangle|\Xi_{mn}\rangle
\nonumber\\
&=&(U^{(mn)}|\Psi\rangle)|\Xi_{mn}\rangle ,
\label{3.9}
\ee
where we have introduced the notation 
\be
\label{3.10}
U^{(mn)}=\sum_{s=0}^{N-1}\exp\frac{-2i \pi sm}{N}
|s-n\rangle\langle s|.
\ee
This result is similar to the one we found in the case of a 
single qubit.  We would now like to examine which transformations
we can perform on the state in the data register by using a
program consisting of a linear combination of the vectors 
$|\Xi_{mn}\rangle$ followed by the action of the processor 
${P}_{123}$ and a subsequent measurement of the program register.

The operators $U^{(mn)}$ satisfy the orthogonality relation
\begin{equation}
\label{orth}
{\rm Tr}\left[ (U^{(m^{\prime}n^{\prime})})^{\dagger}U^{(mn)}
\right] = N\delta_{mm^{\prime}}\delta_{nn^{\prime}} .
\end{equation}
 The space of linear operators ${\cal T}({\cal H})$ 
defined on some Hilbert space $\cal H$
with the scalar product given by (\ref{orth}) we know as
{\it Hilbert-Schmidt space}. Thus the unitary operators $U^{(mn)}$ form
an orthogonal basis in it
and any operator $A\in{\cal T}({\cal H})$ can be expressed in terms of them
\begin{equation}
\label{expans}
A=\sum_{m,n=0}^{N-1}q_{mn}U^{(mn)} .
\end{equation}
The orthogonality relation allows us to find the expansion coefficients
in terms of the operators
\be 
q_{mn}=\frac{1}{N}{\rm Tr}\left[\left(U^{(mn)}\right)^\dagger A
\right].
\ee
Equations (\ref{orth}) and (\ref{expans}) imply that 
\begin{equation}
\sum_{m,n=0}^{N-1}|q_{mn}|^{2}=\frac{1}{N}{\rm Tr}(A^{\dagger}A) .
\end{equation}
Therefore, the program vector that implements the operator $A$ is
given by
\begin{equation}
|v_{A}\rangle_{23} = \left[\frac{N}{{\rm Tr}(A^{\dagger}A)}\right]^{1/2}
\sum_{m,n=0}^{N-1}q_{mn}|\Xi_{mn}\rangle_{23} .
\end{equation}
Application of the processor to the input state $|\Psi\rangle_{1}
|v_{A}\rangle_{23}$ yields the output state
\be
|\Omega\rangle_{123}=
\sum_{mn}q_{mn}U^{(mn)}|\Psi\rangle_{1}\otimes|\Xi_{mn}\rangle_{23}.
\label{3.22}
\ee
To obtain the final result we 
perform a projective measurement of the program register onto vector 
$|M\r_{23}$ 
\be
\label{3.23}
|M\r= \frac{1}{N}\sum_{m,n=0}^{N-1}|\Xi_{mn}\r
\ee 
If the outcome of the measurement is positive, then we 
get the required transformation $A$ acting on an unknown, arbitrary
input state $|\Psi\rangle_1$. 

Let us consider an example.  Suppose we choose for $A$ the unitary 
operator $\openone-2|\phi\rangle\langle\phi |$, where the normalized state 
$|\phi\rangle$ can be expressed as
\begin{equation}
|\phi\rangle = \sum_{k=0}^{N-1}\beta_{k}|k\rangle .
\end{equation}
The expansion coefficients for this operation are given by
\begin{equation}
q_{mn}=\delta_{m0}\delta_{n0}-\frac{2}{N}\sum_{k=0}^{N-1}
e^{2\pi ikm/N}\beta_{k}^{\ast}\beta_{k-n} ,
\end{equation}
and the program vector for this operation is
\begin{equation}
|\Phi\rangle_{23} = |\Xi_{00}\rangle_{23}-\frac{2}{\sqrt{N}}
\sum_{k,n=0}^{N-1}\beta^{\ast}_{-k}\beta_{-(k+n)}|k\rangle_{2}
|k-n\rangle_{3} .
\end{equation}
The program vector can be obtained from a state more closely related
to $|\phi\rangle$ if we introduce a new unitary operator and a 
``complex conjugate'' vector.  Define the operator $W$ by
\begin{equation}
W|k\rangle = |-k\rangle ,
\end{equation} 
and the vector $|\phi^{\ast}\rangle$ by
\begin{equation}
|\phi^{\ast}\rangle = \sum_{k=0}^{N-1}\beta^{\ast}_{k}|k\rangle .
\end{equation}
We then have that
\begin{equation}
\label{simp}
|\Phi\rangle_{23}=(W_{2}\otimes \openone_{3})\left( D_{23}^{\dagger}
\right)^{2}\left(
|\Xi_{00}\rangle_{23}-\frac{2}{\sqrt{N}}|\phi^{\ast}\rangle_{2}
|\phi\rangle_{3}\right) .
\end{equation}
A network that performs the operation $(W_{2}\otimes \openone_{3})
\left( D_{23}^{\dagger}\right)^{2}$ could be added to the input of the 
program register so that the simpler state that appears on the
right-hand side of Eq. (\ref{simp}) could be used as the program.
At the output of the processor we have to perform the projective
measurement discussed in the previous paragraph, and the probability
of achieving the desired result is the same as the probability of
successfully implementing the transformation, $A$.  In this case
the probability is $1/N^{2}$. 

\section{Success Probability}
\label{sec4}
The probability, $p$, of successfully applying the operator $A$ 
to the state $|\Psi\rangle_{1}$ in our example is rather small.
This is because the operator we chose was a linear combination
of all of the operators $U^{(mn)}$.  This means that if
the data register consists of $l$ qubits,
i.e. $N=2^l$, then the probability of a successful implementation 
of a general transformation $A$ decreases exponentially with the 
size of the data register.  However, if we were to choose an
operator, or set of operators, that was a linear 
combination of only a few of the $U^{(mn)}$, then the success
probability can be significantly improved.  This would entail
making a different measurement at the output of the program 
register.  Instead of making a projective measurement onto the
vector $|M\rangle$, one would instead make a measurement onto the
vector
\be
|M^{\prime}\r= \frac{1}{{\cal N}^{1/2}}\sum_{m,n:q_{mn}\ne 0}
|\Xi_{mn}\r
\ee
where ${\cal N}$ is the total number of nonzero coefficients $q_{mn}$,
in the decomposition in Eq. (\ref{expans}).
If the operation being implemented is unitary, then, in this case,
the probability of implementing it is 
\be
\label{4.1}
p= \frac{1}{\cal N} .
\ee 
where ${\cal N}$ is the total number of nonzero coefficients $q_{mn}$,
in the decomposition (\ref{expans}).
There are, in fact, large classes of operations  
that can be expressed in terms of a small number of operators 
$U^{(mn)}$ \cite{footnote1}. For these operators,
the probability of success can be relatively large and, in
principle, independent of the size of the Hilbert space of the data
register. 

{\em Example 1.}\newline
Let us consider the one-parameter set of unitary
transformations $U_\varphi$
\be
\label{4.2}
U_\varphi
= \cos\varphi \, \openone
+i\sin\varphi\, 
\left[\frac{1+i}{2} U^{(01)}
+\frac{1-i}{2} U^{(03)}\right]
\ee 
where the unitaries $U^{(mn)}$ are given by Eq.(\ref{3.10}).
These unitaries for $N=4$ can be explicitly written as
\be
U^{(01)} = \sum_{s=0}^3 (-i)^s P_s \, ;
\qquad
U^{(03)} = \sum_{s=0}^3 (i)^s P_s \, ,
\label{4.3}
\ee
where $P_s=|s\r\l s|$.
From here we find the  expression for the operator
(\ref{4.2}) in the form:
\be
\label{4.4}
U_\varphi
= \cos\varphi \, \openone
+i\sin\varphi\, 
\left[P_0+P_1-P_2-P_3\right] \, ,
\ee 
We note that if we rewrite the parameters $s$ as binary numbers,
$s=j_{1}2+j_{0}$, where $j_{k}$ is either $0$ or $1$, and express 
the states $|s\rangle$ as tensor products of qubits, i.e.\ 
$|s\rangle =|j_{1}\rangle\otimes |j_{0}\rangle$,
we find that the operator in brackets on the right-hand side of
Eq.~(\ref{4.2}) can be expressed as
\be
\label{rel}
\left[\frac{1+i}{2} U^{(01)}
+\frac{1-i}{2} U^{(03)}\right]  = \sigma_3 \otimes \openone.
\ee
From Eq.\ (\ref{4.4}) it is clear that $U_\varphi$ has eigenvalues
of magnitude $1$, which implies that $U_\varphi$ is unitary.
It can be realized by the universal quantum
processor (\ref{3.5}) with a 
probability of successful implementation equal to $1/3$.
This example illustrates that it is possible to realize
large classes of unitary operations with a probability that is
greater than the reciprocal of the dimension of the program
register.

This example can be easily generalized.  Consider a one-parameter 
set of unitary operators acting on a Hilbert space consisting of 
$l$ qubits, which is given by 
\be
\label{4.10}
U_\varphi= \cos\varphi \, \openone^{\otimes l} +i\sin\varphi\,
\sigma_3 \otimes\openone^{\otimes (l-1)} 
\ee
The operator $\sigma_3\otimes\openone^{\otimes(l-1)}$ 
is diagonal and therefore only the diagonal unitaries from our 
set $U^{(mn)}$, i.e. $U^{(m0)}$, appear in its expansion, 
Eq.\ (\ref{expans}) . 
Moreover the coefficients $q_{m0}$ in the expansion
are non-vanishing only for odd $m$. It follows that 
\be
U_\varphi= \cos\varphi \, \openone^{\otimes l} +i\sin\varphi\,
\sum_{{\rm odd}\ m} q_{m0} U^{(m0)} ,
\ee
and the probability of a successful implementation of this unitary 
transformation is $p=2/(2^l+2)$.

{\em Example 2.}\newline
For some sets of operators it is possible to do even better than
we were able to do in the previous example.
Consider the one-parameter set of unitary operators 
given by
\begin{equation}
\label{ex2}
U_\vartheta=\cos\vartheta \, \openone+ i\sin\vartheta \, U^{(0,N/2)} ,
\end{equation}
where $N$ is assumed to be even.  
That this operator is unitary
follows from the fact that $U^{(0,N/2)}$ is self-adjoint.  A
program vector that would implement this operator is
\begin{equation}
|\Phi\rangle_{23}=\cos\vartheta |\Xi_{00}\rangle_{23}+i\sin\vartheta
|\Xi_{0,N/2}\rangle_{23} ,
\end{equation}
and at the output of the program register we make a projective
measurement corresponding to the vector
\begin{equation}
|M\rangle_{23}=\frac{1}{\sqrt{2}}(|\Xi_{00}\rangle_{23}
+|\Xi_{0,N/2}\rangle_{23} .
\end{equation}
The probability for successfully achieving the desired result,
i.e.\ the vector $U_\vartheta |\Psi\rangle_{1}$ in the data register,
is $1/2$ irrespective of the value $N$, i.e. the number of qubits.

\section{Conclusion}
We have presented here a programmable quantum processor that exactly
implements a set of operators that form a basis for the space
of operators on qudits.  This processor has a particularly simple
representation in terms of elementary quantum gates.  It is, however,
by no means unique.  It is possible, in principle, to build a processor
that exactly implements any set of unitary operators that form a
basis for the set of operators on qudits of dimension $N$, and uses any orthonormal set of $N^{2}$ vectors as programs.  Explicitly, if the
set of operators is $\{ V_{n}|n=1,\ldots N^{2}\}$ and the program
vectors are $\{|y_{n}\rangle |n=1,\ldots N^{2}\}$, the processor
transformation is given by
\begin{equation}
P_{dp}=\sum_{n=1}^{N^{2}}V_{n}^{(d)}\otimes |y_{n}\rangle_{p}
\,_{p}\langle y_{n}| ,
\end{equation}
where the superscript $(d)$ on the operator $V_{n}$ indicates 
that it acts on the data register.  

As an example, consider a data register consisting of $l$ qubits.
We could use the processor discussed in section III to perform
operations on states in this register, but we can also do
something else; we can use $l$ single-qubit processors, one for each
qubit of the data register. Specifically, our unitary basis for 
the set operations on the data register would be 
\be
U_{JK}=U_{j_1k_1,\dots,j_lk_l}=\bigotimes_{m=1}^l S_{j_{m}k_{m}}
\label{4.8}
\ee
where $J=(j_1,\dots,j_l)$ and $K=(k_1\dots,k_l)$ are sequences of 
zeros and ones, and the operators $S_{j_{m}k_{m}}$ are the defined
immediately 
after Eq.\ (\ref{trans}).  The program register would consist of 
$l$ pairs of qubits, $2l$ qubits in all, with each pair controlling 
the operation on one of the qubits in the data register.  Each of the operators in our basis can be implemented perfectly by a program 
consisting of the tensor product state, 
$\prod_{m=0}^{l}|\Xi_{j_{m}k_{m}}^{(m)}\rangle$, where $|\Xi_{j_{m}k_{m}}^{(m)}\rangle$ is a two-qubit state that implements
the operation $S_{j_{m}k_{m}}$ on the $m$th qubit of the data register.

We are then faced with the problem of which processor to use.  This
very much depends on the set of operations we want to apply to the
data.   How to choose the processor so that a given set of operations 
can be implemented with the greatest probability, for a fixed size of
the program register is an open problem.  A second issue is simplicity.
One would like the processor itself and the program states it uses to
be as simple as possible.  The simplicity of the processor is related
to the number of quantum gates it takes to construct it.  We would 
maintain that the processors we have presented here are simple, though
whether there are simpler ones we do not know.  Judging the simplicity
of the program states is somewhat more difficult, but they should be 
related in a relatively straightforward way to the operation that they
encode.  In many cases these states will have been produced by a 
previous part of a quantum algorithm, and complicated program states
will mean more complexity for the algorithm that produces them.  The
program states proposed by Vidal and Cirac and the ones proposed by 
us in section II are, in our opinion, simple.

A final open problem that we shall mention, is finding a systematic 
way of increasing the probability of successfully carrying out a
set of operations by increasing the dimensionality of the space
of program vectors.  Vidal and Cirac showed how to do this in
a particular case, but more general constructions would be 
desirable \cite{Vidal2000}.  Doing so would give one a method
of designing programs for a quantum computer.

\acknowledgements
This work was supported in part
by the European Union  projects EQUIP and QUEST, 
 and by the National Science Foundation under grant
PHY-9970507.

\end{multicols}
                       

\begin{thebibliography}{99}
\bibitem{footnote1}
We note that an arbitary sum of unitary 
operators $U^{(mn)}$ is not necessarily a unitary
operator.
\end{thebibliography}
\end{document}